\documentclass[11pt]{revtex4-2}
\usepackage{relsize}
\usepackage{hyperref}
\usepackage{amsmath,amsfonts,amssymb}
\hypersetup{dvips,dvipdfm,colorlinks=true,urlcolor=magenta,filecolor=magenta,linktoc=page,citecolor=red,linkcolor=blue,bookmarks=true}
\usepackage{graphicx,epstopdf}
\begin{document}
	
	\title{Does diffusion mechanism favour the emergent scenario of the universe?}
	\author{Subhayan Maity\footnote {maitysubhayan@gmail.com}}
	\affiliation{Department of Mathematics, Jadavpur University, Kolkata-700032, West Bengal, India.}

	\author{Subenoy Chakraborty\footnote {schakraborty.math@gmail.com}}
	\affiliation{Department of Mathematics, Jadavpur University, Kolkata-700032, West Bengal, India.}

\begin{abstract}
 In the present work, the flat FLRW Universe has been modelled with  cosmic matter in the form of  diffusive barotropic fluid. The diffusive fluid undergoes dissipation due to diffusion mechanism  in the form of cosmological scalar field $\phi$.  From the perspective of non-equilibrium thermodynamics, the evolution equations of the universe have been formulated. By a suitable choice of the cosmological scalar field, emergent scenario of the universe has been obtained.

\end{abstract}

\maketitle





Diffusion can be considered as one of the basic macroscopic forces in nature. Several physical and biological processes are caused due to diffusion. Some well known examples of dynamical processes (in physics) are heat conduction, Brownian motion and various transport phenomena \cite{Franchi:10485F,Haba:2009by,Herrmann:024026,Calogero:2011re} in biological systems where diffusion is the driving mechanism . The random collisions between the particles of the system and those of the background is caused due to diffusion mechanism at the microscopic level. On the other hand, random effects are averaged at the macroscopic scale and diffusion is characterised by heat equation or Fokker-Planck equation . Although there is a wide variety of phenomena having diffusive behaviour, still there does not exist a consistent diffusion theory in general relativity. However from cosmological point of view, it is speculated that diffusion may have a basic role in the evolution dynamics of the large scale structure formation of the universe. Further, in standard cosmology, galaxies are assumed as point particles of a fluid, undergoing velocity diffusion \cite{Franchi:10485F,Haba:2009by,Herrmann:024026,Calogero:2011re,Calogero:2012kd}. 

\par To consider diffusion in general relativity, one has to consider macroscopic continuum description provided by the Fokker-planck equation. So, in diffusive process, the energy - momentum tensor is not covariantly conserved (i.e. $\nabla_{\mu}  T^{\mu \nu}\neq 0$), rather it satisfies the Fokker - Planck equation, namely 
   \cite{Calogero:2011re,Calogero:2012kd,Franchi:10485F,Haba:2009by,Herrmann:024026}
\begin{equation}
	\nabla _{\mu} T^{\mu \nu}=3 \sigma J^{\nu},   \label{1}
\end{equation} 
where $\sigma (> 0)$ is the diffusion constant and $J^{\nu}$ , the current density of the matter satisfies
\begin{equation}
	\nabla _{\mu} J^{\mu}=0 .  \label{2}
\end{equation}
Thus one can not have usual Einstein equations i.e. $R_{\mu \nu}-\dfrac{1}{2}R g_{\mu \nu}= T _{\mu \nu} $ for diffusive process, due to Bianchi identity. The simplest modification of the Einstein equation is to introduce two interacting matter components of which one is the usual diffusive fluid having conservation (non-conservation) equation given by equation (\ref{1}) while the simplest choice for the other component (in analogy with cosmological constant) is a cosmological scalar field. so the modified Einstein field equations take the form 
\begin{equation}
	 R_{\mu \nu}-\frac{1}{2}R g_{\mu \nu}+\phi g _{\mu \nu}= T _{\mu \nu}   ,\label{3}
\end{equation}
where the scalar field $\phi$ has the evolution equation \cite{Calogero:2011re,Franchi:10485F,Haba:2009by,Herrmann:024026} (dimension factor in $\phi$ has been chosen to be unity for convenience.)
\begin{equation}
	\nabla _{\mu} \phi=3 \sigma J _{\mu}  ,  \label{4}
\end{equation}
and $T_{\mu \nu}$ satisfies the above Fokker - Planck equation (given by equation (\ref{1})).

 Here $3 \sigma $ measures the energy transferred from the scalar field to the matter per unit time due to diffusion. Note that in vacuum or in the absence of diffusion, the above modified Einstein field equations (\ref{3}) become Einstein equations with a cosmological constant while in general equation (\ref{3}) may be termed as Einstein equations with variable `cosmological constant'. 
 \par The above diffusion process is usually termed as kinetic model with microscopic velocity of the fluid particles undergoing diffusion. Here the diffusion mechanism takes place on the tangent bundle of the space time and as a result Lorentz invariance of the space-time is preserved. Now choosing the cosmic fluid as perfect fluid, one has
 the energy-momentum tensor 
 \begin{equation}
 	\mathcal{T}_{\mu \nu} =\rho u_{\mu}u_{\nu}+p(g_{\mu \mu}+u_{\mu}u_{\nu})     \label{5}
 \end{equation}
with current density $J^{\mu}=n u^{\mu}$. Here $n$ is the particle number density of the fluid. $u^{\mu}$ is the four velocity of the fluid. $\rho$ and $p$ are the energy density and thermodynamic pressure of the fluid respectively.
Now projecting equation (\ref{1}) along the fluid $4$-velocity  $u^{\mu}$ and on the hyper surface orthogonal to $u^{\mu}$ , one gets \cite{Calogero:2011re,Franchi:10485F,Haba:2009by,Herrmann:024026}
\begin{equation}
	\nabla_{\mu} (\rho u^{\mu}) +p \nabla _{\mu} u ^{\mu} =3\sigma . n   \label{6}
\end{equation}  and 
\begin{equation}
	(p+\rho)u^{\mu}\nabla_{\mu} u ^{\nu}+ (u ^{\mu} u^{\nu}+ g^{\mu \nu})\nabla _{\mu} p= 0 ,  \label{7} 
\end{equation}  
Here equation (\ref{7}), the Euler equation does not change due to diffusion process as diffusion force  acts along the matter flow. 
\textbf{It is to be noted that there are several diffusion models in the literature namely for unification of dark energy and dark matter from diffusive cosmology see ref. \cite{Benisty:2018oyy}. Ref. \cite{Benisty:2017lmt} deals with transition  between bouncing hyper-inflation to $\Lambda_{CDM}$ from diffusive scalar fields while unified DE-DM with diffusive interactions and interacting diffusive unified dark energy and dark matter from scalar fields can be found in ref.\cite{Benisty:2017rbw} and \cite{Benisty:2017eqh} respectively. In particular, a Lagrangian formulation of diffusion mechanism can be found in ref. \cite{Benisty:2018oyy}.} 

   In the background of homogeneous and isotropic flat FLRW model, the modified Friedmann equations with diffusion dynamics take the form, 
   \begin{equation}
   	3 H^2= \rho +\phi     \label{8}
   \end{equation}  and
\begin{equation}
	2 \dot{H}=-(\rho +p)      \label{9}
\end{equation}
 Now equation (\ref{3}) for the present geometry simplifies to 
 \begin{equation}
 	na^3(t)=\mbox{constant, i.e.~} n=n_0a^{-3} .  \label{10}
 \end{equation}
 Hence the modified matter conservation equation (\ref{1}) for the matter field (\ref{5}) takes the form,
 \begin{equation}
 	\dot{\rho}+3H(p+\rho)=\sigma n_0 a^{-3}=\sigma_0 a^{-3}   \label{11}
 \end{equation} which on integration yields 
\begin{equation}
	\rho =a^{-3(1+\omega)} \left [\rho _0 +\int\limits_{t_0}^{t}\sigma _0 a^{3 \omega } dt\right ]. \label{12}
\end{equation}
 Here $\omega =\dfrac{p}{\rho}$, is the constant equation of state parameter of the fluid. $\rho _0$ is the energy density at reference epoch of time $t=t_0$ and  $a(t_0)=1$. $n(t_0)=n_0$ is assumed. Now eliminating $\rho$ between equations (\ref{8}) and (\ref{9}), one gets the cosmic evolution equation as
 \begin{equation}
 	2 \dot{H}+3(1+\omega)H^2 =\phi (1+\omega)   \label{13}
 \end{equation} On the other hand, the above modified Friedmann equations (i.e. equations (\ref{8}) and (\ref{9})) for diffusive mechanism can be rewritten as,
 \begin{equation}
 	3H^2 =\rho _d   ~\mbox{,}~ 2 \dot{H}= -(\rho _d +p_d + \pi _d)  \label{14}
 \end{equation} while the conservation equation(\ref{11}) becomes
\begin{equation}
	\dot{\rho _d}+ 3 H(\rho _d +p_d + \pi _d)=0 ,   \label{15}
\end{equation}
 with $\rho _d =\rho +\phi$, $p _d =p$ and $\pi _d =-\phi$ . \textbf{Thus interacting two fluid system in diffusion mechanism \cite{Calogero:2013zba} is equivalent to a single dissipative fluid in Einstein gravity.} Here dissipation is chosen in the form of bulk viscous pressure $\pi_d$. Further one may consider the above dissipative pressure (i.e. bulk viscous pressure)   due to non-equilibrium thermodynamics with particle creation mechanism. In fact, for adiabatic thermodynamic process,  the dissipative pressure $\pi _d$
 is related linearly to the particle creation rate $\Gamma _d$ as \cite{Chakraborty:2014ora,Chakraborty:2014oya}
\begin{equation}
	\pi _d =-\frac{\Gamma_d}{3 H}(\rho _d +p_d) .  \label{16}
\end{equation}       Using the $1$st friedmann equation in (\ref{14}) of equivalent Einstein gravity  into the above equation (\ref{16}) with
$p_d=p=\omega \rho$ and $\pi _d =-\phi$, the cosmological scalar field is  related to the particle creation rate as
\begin{equation}
	\Gamma _d  =\frac{3 H \phi}{3 H^2(1+\omega)-\omega \phi}  . \label{17}
\end{equation}   
 \textbf{Hence the present interacting diffusive mechanism \cite{Calogero:2013zba} with cosmological scalar field can be considered as non-equilibrium thermodynamic description of Einstein gravity with particle creation formalism.}
 \par \textbf{To overcome the classical singularity of Einstein gravity, cosmologists propose two models namely the bouncing Universe or the emergent Universe. In the present work, for non-singular solution we shall consider the model of emergent scenario as it is very much relevant as pre-inflationary era. An emergent Universe \cite{Ellis:2003qz,Ellis:2002we,Guendelman:2014bva,Bose:2020xml,Chakraborty:2014ora,Bhattacharya:2016env,Banerjee:2007qi}
 is a modelled Universe with no time like singularity having static Einstein era in the infinite past (i.e. $t\rightarrow -\infty$).  The present work examines whether emergent scenario is possible or not in the present cosmological scalar field diffusion mechanism. In order to have a cosmological solution one may choose phenomenologically the form of $\phi$ as}
\begin{equation}
	\phi=3 \alpha H ,   \label{18}
\end{equation}
with $\alpha $, a constant.
Using this choice of $\phi$ in the field equations (\ref{8}),(\ref{9})and (\ref{11}) one gets 
\begin{equation}
	\sigma_0 a^{-3} =-3 \alpha \dot{H},    \label{19}
\end{equation}
\textbf{which shows that $\alpha $ and $\sigma _0$  are of same sign (due to $\dot{H}<0$).}

 For this choice of $\phi$, the solutions of the cosmic evolution equation (\ref{13}) yield the form of Hubble parameter and scale factor as,
\par(i)~~For $\alpha \geq H_0$ :
\begin{equation}
	H=\frac{\alpha}{1+\left (\frac{\alpha}{H_0}-1 \right )e^{-\frac{3}{2}\alpha(1+\omega)(t-t_0)}}  ~~,      \label{20}
\end{equation} 
\begin{equation}
	a=\left[\frac{\left (\frac{\alpha}{H_0}-1 \right )+e^{\frac{3}{2}\alpha(1+\omega )(t-t_0)}}{\left (\frac{\alpha}{H_0}-1 \right )+1}\right]^{\frac{2}{3(1+\omega)}}         \label{21}
\end{equation}
\par (ii)~~For $0< \alpha < H_0$ :
\begin{equation}
	H=\frac{\alpha}{1-\left (1-\frac{\alpha}{H_0}\right )e^{-\frac{3}{2}\alpha(1+\omega)(t-t_0)}}  ~~,   \label{22}
\end{equation}

\begin{equation}
	a=\left[\frac{\left (1-\frac{\alpha}{H_0}\right )-e^{\frac{3}{2}\alpha(1+\omega )(t-t_0)}}{\left (1-\frac{\alpha}{H_0}\right )-1}\right]^{\frac{2}{3(1+\omega)}}          \label{23}
\end{equation}  \par and \par (iii)~~For $\alpha < 0$ :
\begin{eqnarray}
	H=\frac{|\alpha|}{1-\left (\frac{|\alpha|}{H_0}+1\right )e^{-\frac{3}{2}\alpha(1+\omega)(t-t_0)}}  \label{24} ~~\mbox{,}~~~ \\    
	a=\left[\frac{e^{\frac{3}{2}\alpha(1+\omega )(t-t_0)}-\left (\frac{|\alpha|}{H_0}+1 \right )}{1-\left (\frac{|\alpha|}{H_0}+1 \right)}\right]^{\frac{2}{3(1+\omega)}}.      \label{25}
\end{eqnarray}
Here $H_0$ is the value of $H$ at reference epoch of time $t_0$ .
\par For $\alpha<0$, the above cosmological solution (\ref{25}) has a big-bang singularity at the epoch,
\begin{equation}
	t_s=t_0-\frac{2}{3(1+\omega )|\alpha |}\ln \left (1+\frac{|\alpha|}{H_0}\right)  .    \label{26}
\end{equation} 

\textbf{Note that $\alpha <0$ (i.e. $\sigma_0 <0$) is not physically realistic, so we shall present the above solution for $\alpha <0$ only for mathematical completeness.}

\par Again for  $0< \alpha < H_0 $,  big-rip singularity exists for the cosmological solution  (\ref{23}) at the epoch,
\begin{equation}
	t_s=t_0+\frac{2}{3(1+\omega)\alpha}\ln\left[\left(1-\frac{\alpha }{H_0}\right)\right]  .  \label{27}
\end{equation} 
\par  In the case $ H_0<\alpha $, the cosmological solution (\ref{21}) has no singularity at any real time.

\par Clearly, this solution [(\ref{20}) , (\ref{21})] yields the Emergent scenario as it follows the  following criteria \cite{Chakraborty:2014ora} :
\begin{subequations}
	\begin{align}
		H &\rightarrow 0~, ~a \rightarrow \left[\frac{\alpha-H_0}{\alpha}\right]^{\frac{2}{3(1+\omega)}} \mbox{when}~ t\rightarrow -\infty    \label{28a}\\
		H &\rightarrow 0~,~ a \rightarrow \left[\frac{\alpha-H_0}{\alpha}\right]^{\frac{2}{3(1+\omega)}} \mbox{when}~ t<<t_0 ~\mbox{and}~   \label{28b}\\
		H &\sim \alpha~,~ a\simeq \left[\frac{H_0}{\alpha}\right]^{\frac{2}{3(1+\omega)}} \exp \left [{\alpha(t-t_0)}\right ] ~~\mbox{when}~  t>>t_0     \label{28c}
	\end{align} 
\end{subequations}

\par So  evidently the explicit solution for emergent scenario should be in the form (also considering , $\alpha =H_0+\delta $ with $\delta \geq 0$) :
\begin{subequations}
	\begin{align}
		H^{(E)}=\frac{(H_0+\delta) H_0}{H_0+\delta  e^{-\frac{3}{2}(H_0+\delta)(1+\omega)(t-t_0)}}  \label{29a}      \\
		\mbox{and}~
		a^{(E)}=\left[\frac{\delta+H_0 e^{\frac{3}{2}(H_0+\delta)(1+\omega )(t-t_0)}}{H_0+\delta}\right]^{\frac{2}{3(1+\omega)}}         \label{29b}
	\end{align}
\end{subequations} 
under the diffusive  non-singular scalar field ,
\begin{equation}
	\phi=3 (H_0+\delta)H .  \label{30}
\end{equation}
The nature of corresponding particle creation rate can be found from equation (\ref{17}) as,
\begin{equation}
	\Gamma _d =\frac{3H}{1-(1+\omega)(1-\frac{H}{\alpha})} .  \label{31}
\end{equation}

\begin{figure}[h]

	\begin{minipage}{0.47\textwidth}
		\centering
		\includegraphics*[width=0.9\linewidth]{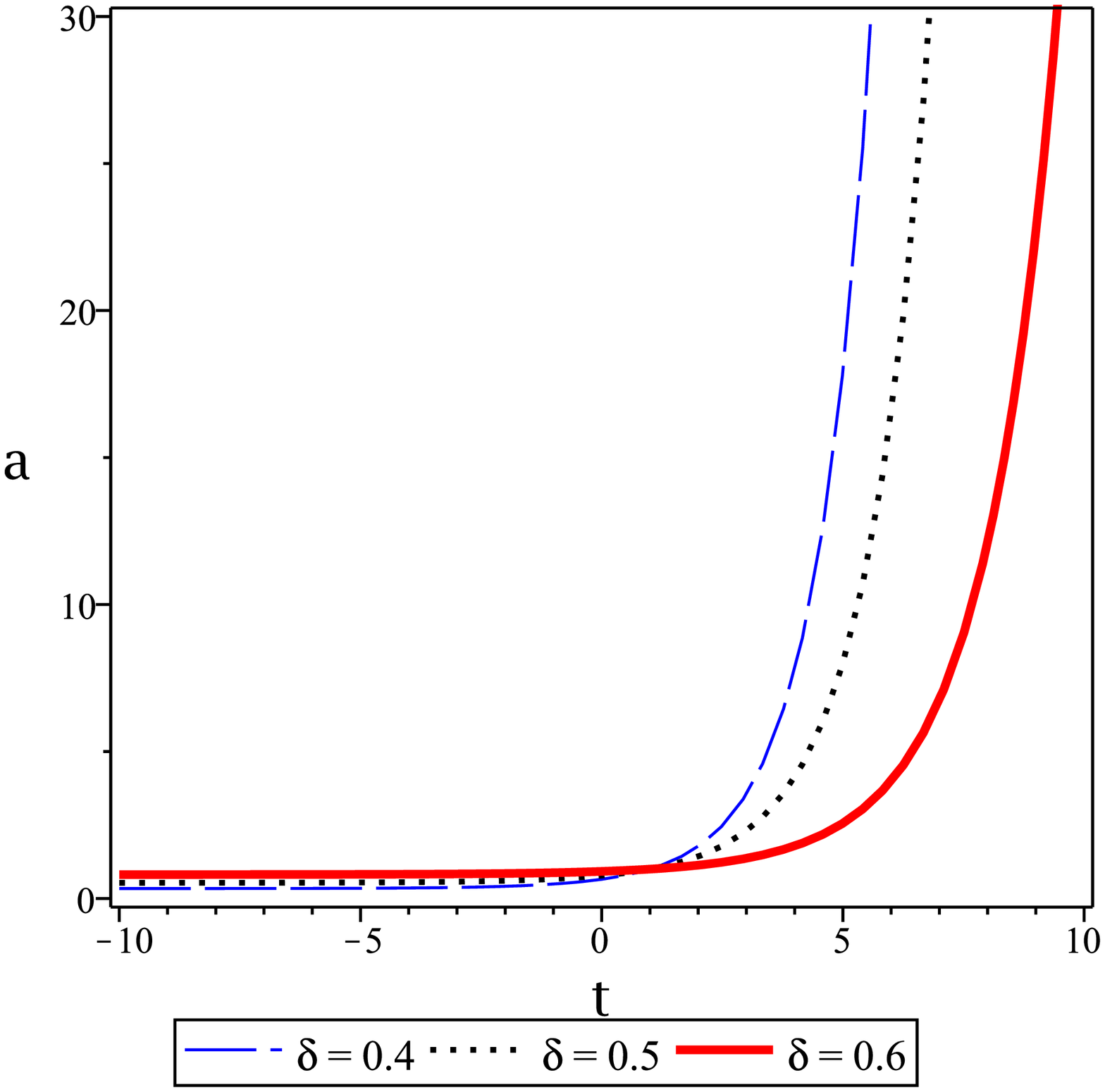}\\
		(a) 
	\end{minipage}
	\begin{minipage}{0.47\textwidth}
		\centering
		\includegraphics*[width=0.9\linewidth]{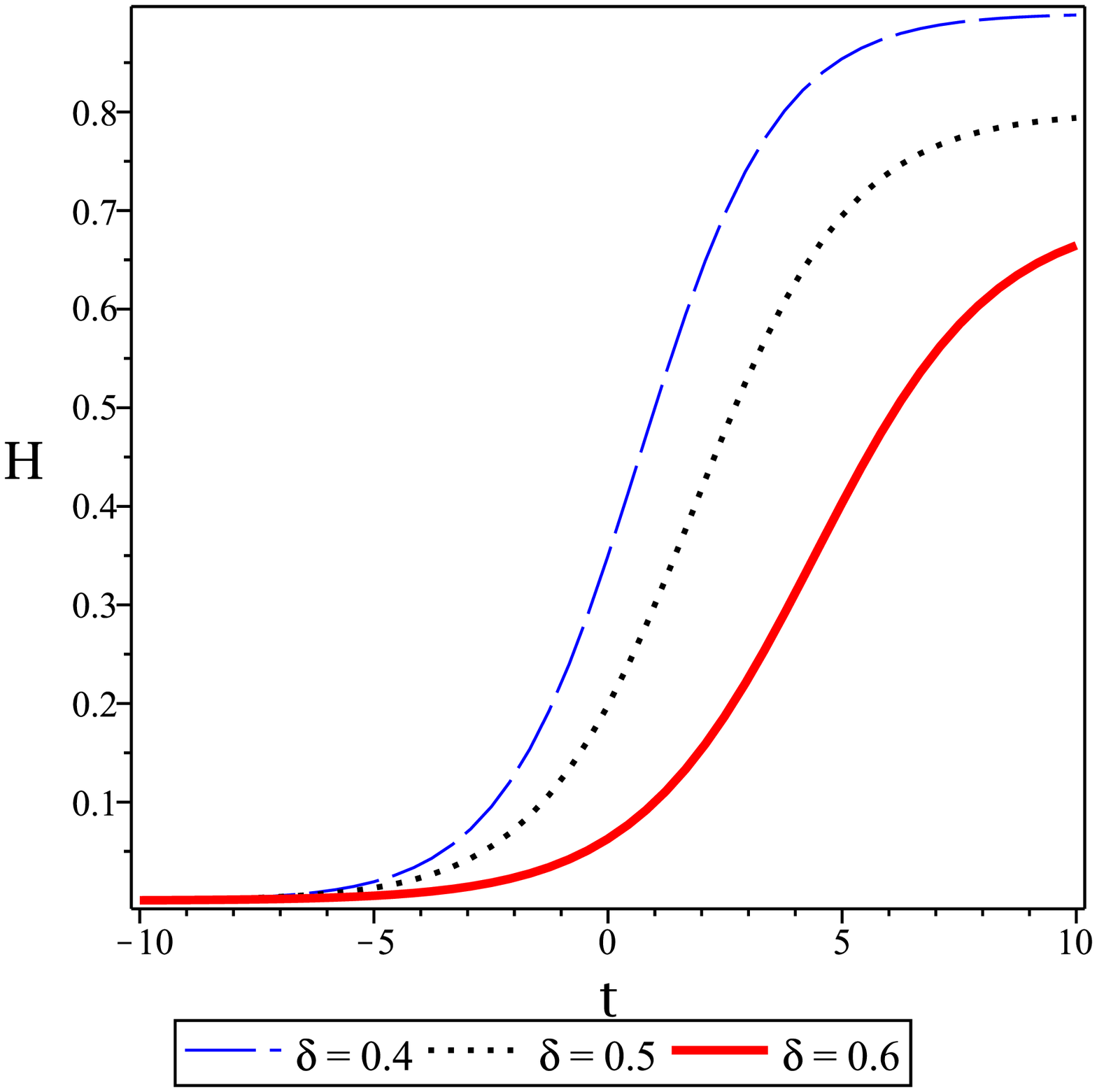}\\
		(b)
		
	\end{minipage}
	\begin{minipage}{0.47\textwidth}
		\centering
		\includegraphics*[width=0.9\linewidth]{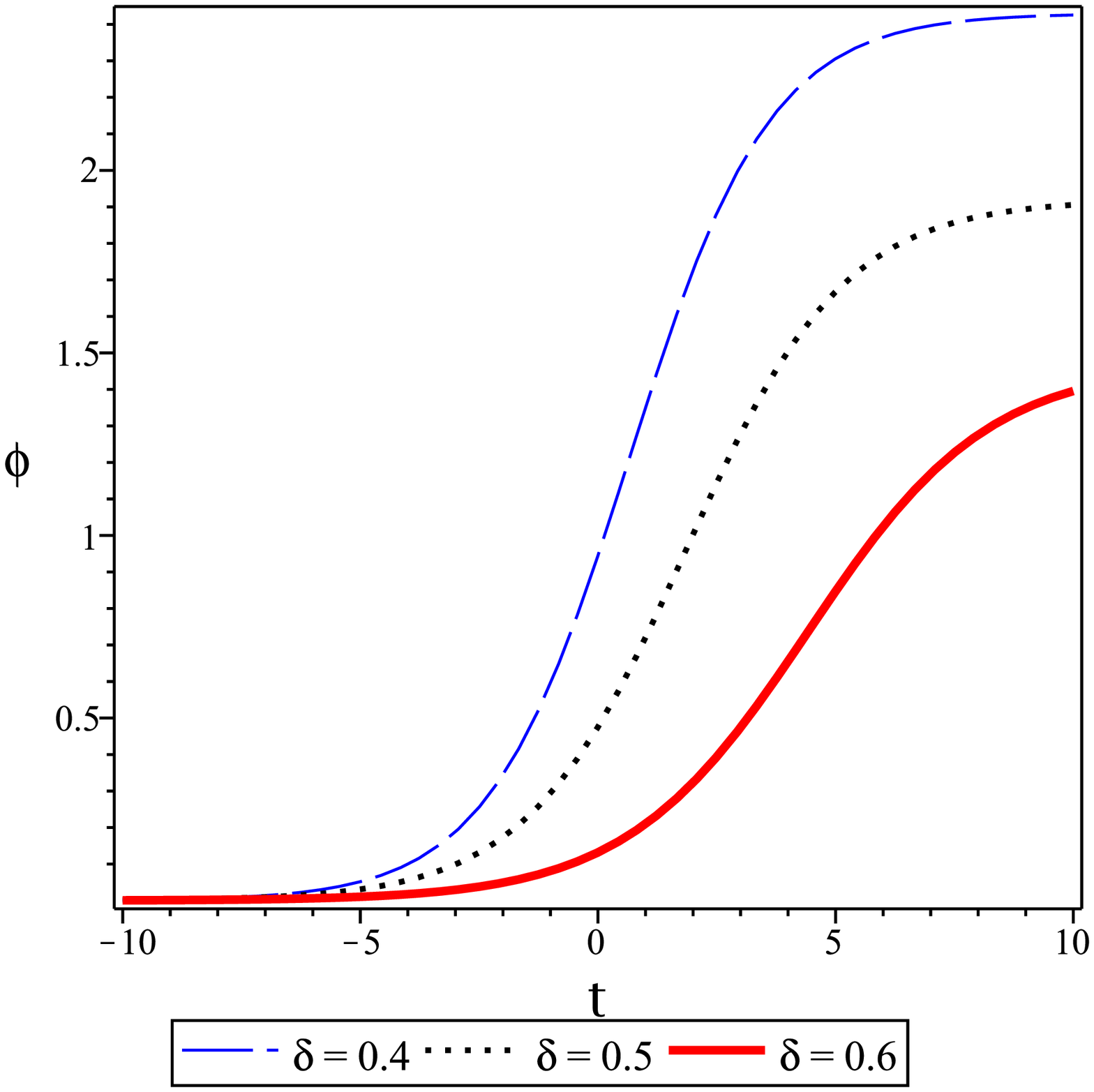}\\
		(c) 
		
	\end{minipage}
	\begin{minipage}{0.47\textwidth}
		\centering
		\includegraphics*[width=0.9\linewidth]{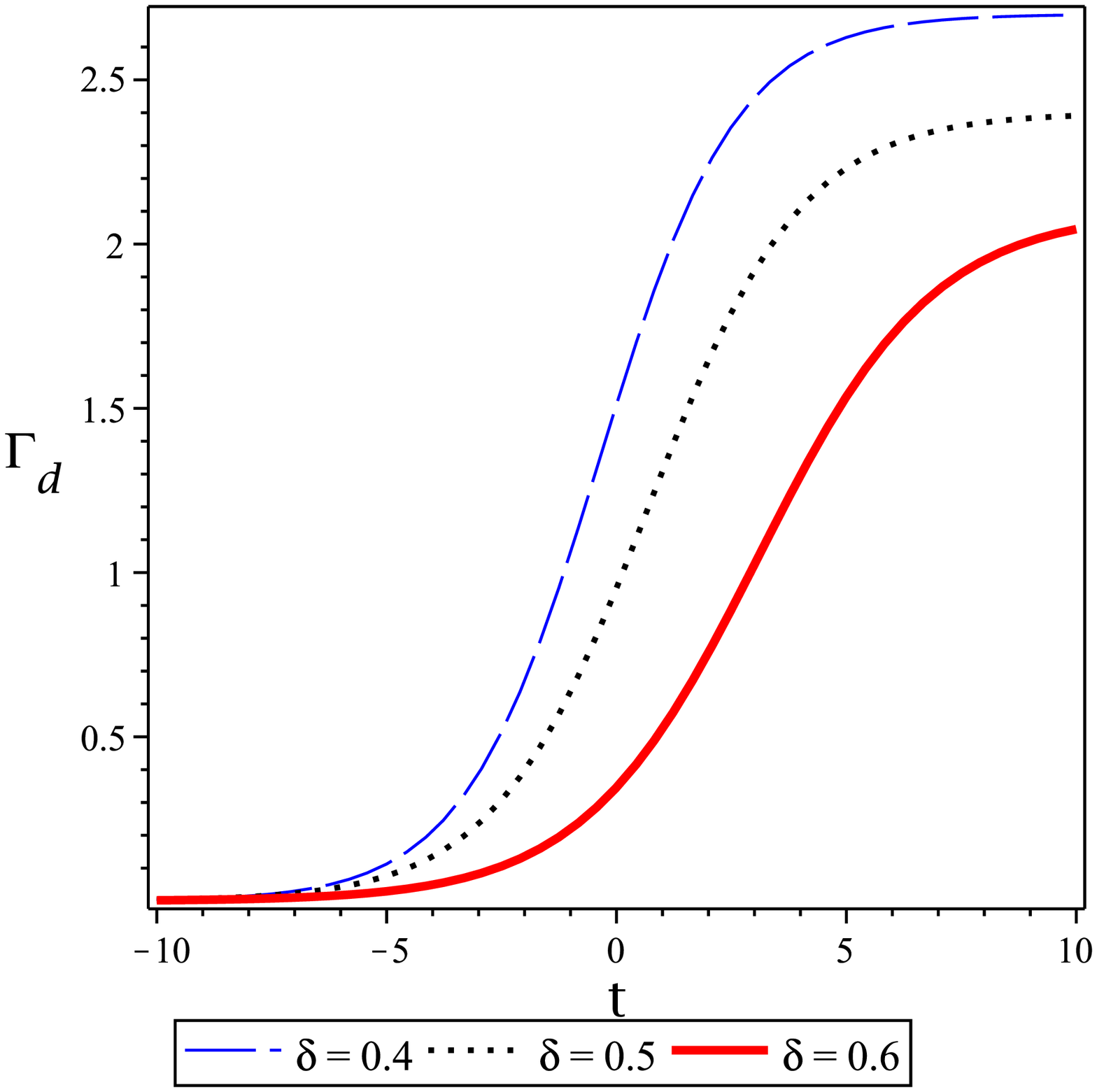}\\
		(d) 
	\end{minipage}
	
	\begin{minipage}{0.87\textwidth}
		\caption{ Evolution of different  physical parameters namely (a) Scale factor $a$ (top left), (b) Hubble parameter $H$  (top right), (c) Cosmological scalar field $\phi$  (bottom left), (d) Particle creation rate $\Gamma_d$ (bottom right)  for $\omega =-0.5$, $\alpha=0.9$, $t_0=1$ with three different values of $H_0$ : $0.5$, $0.4$, $0.3$. }.\label{fig1}
	\end{minipage}
\end{figure}

\begin{figure}[h]
	
	\begin{minipage}{0.47\textwidth}
		\centering
		\includegraphics*[width=0.9\linewidth]{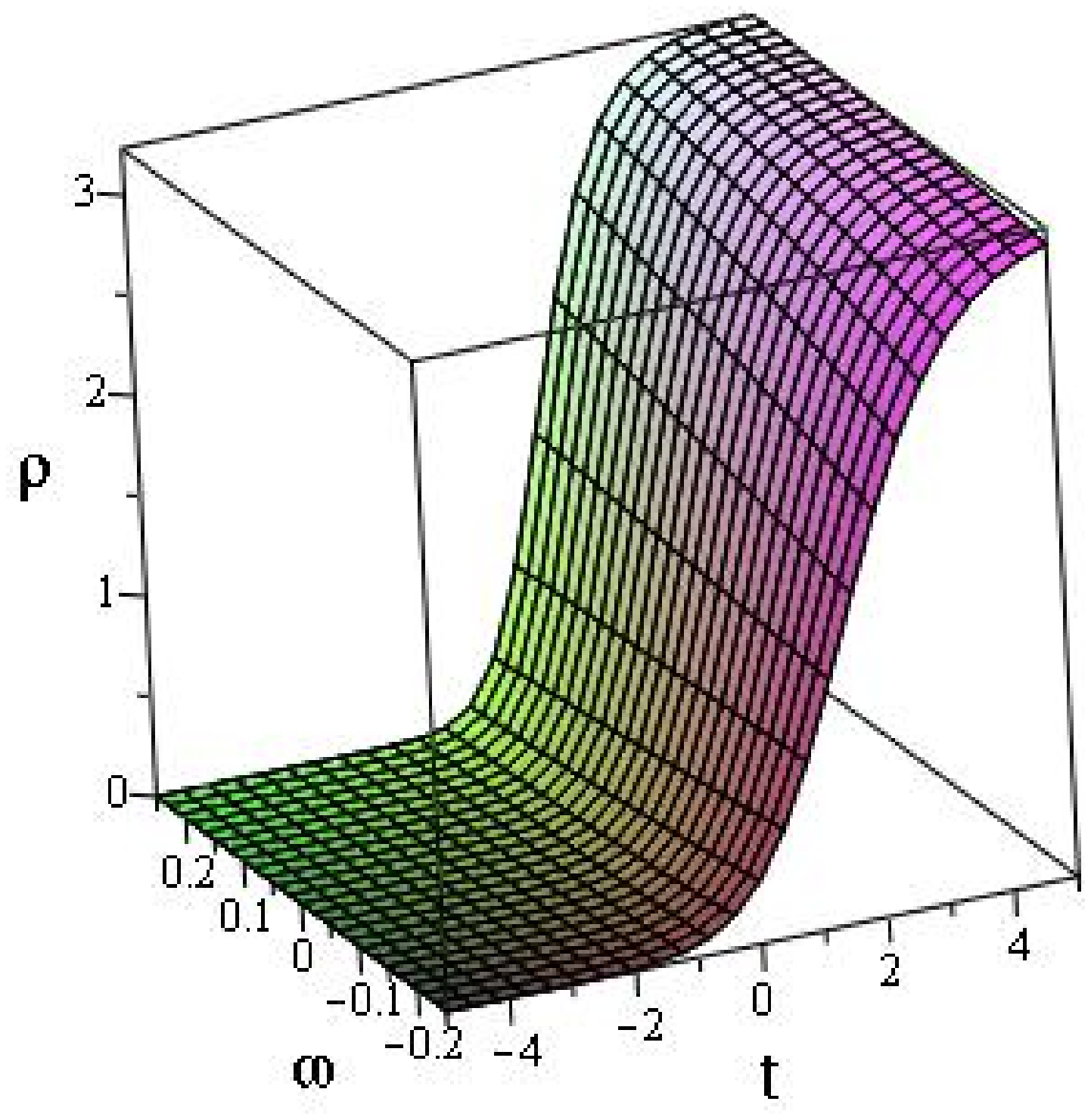}\\
		(a) 
	\end{minipage}
	\begin{minipage}{0.47\textwidth}
		\centering
		\includegraphics*[width=0.9\linewidth]{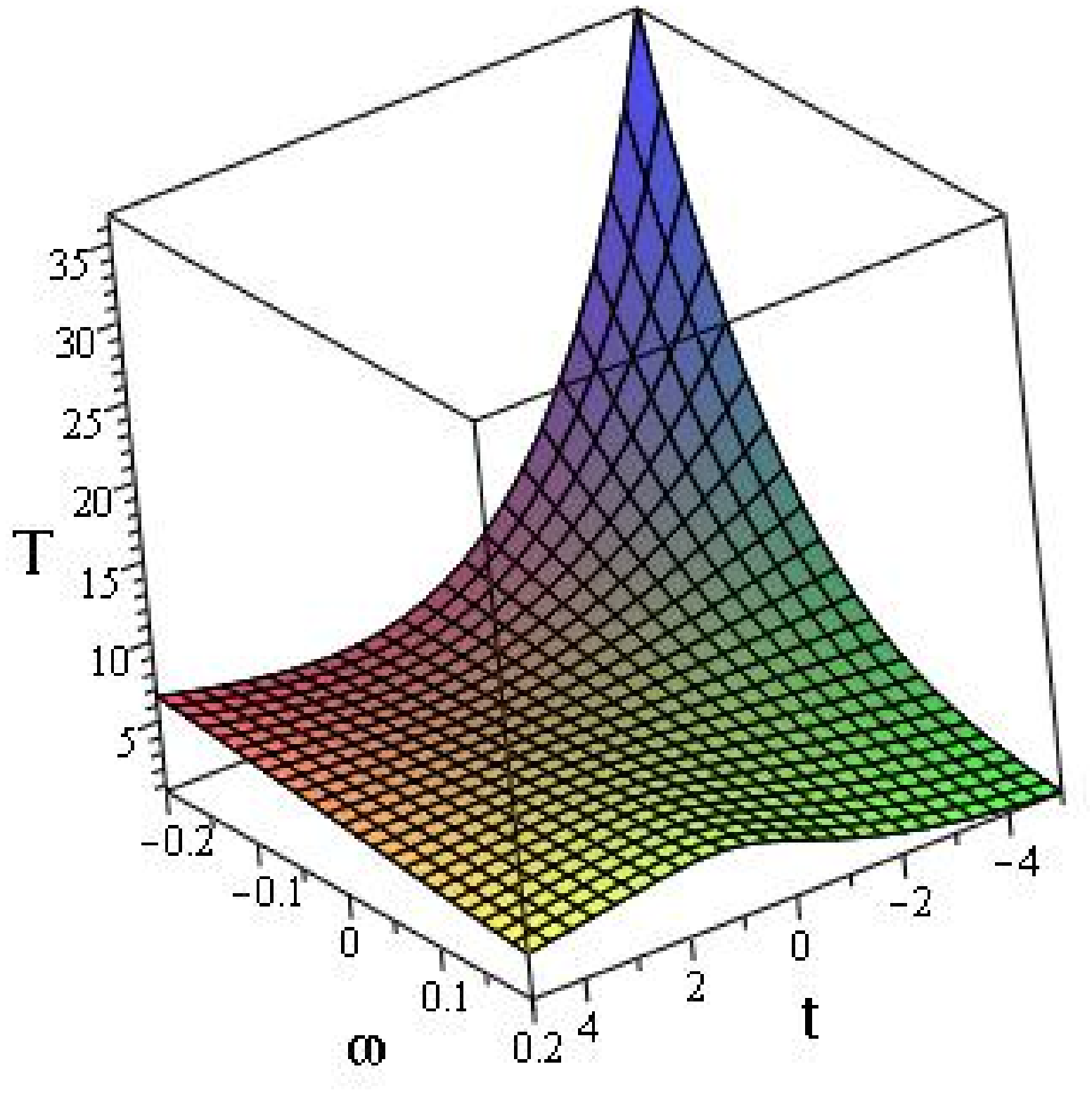}\\
		(b)
		
	\end{minipage}
	
	\begin{minipage}{0.76\textwidth}
		\caption{ Evolution of different  thermodynamic parameters namely (a) Energy density  $\rho$ (left) and (b) Temperature $T$  (right) as a functions of time $t$ and barotropic index of the fluid $\omega$ for  $H_0=0.5,t_0=1$ with  $\delta=0.4$.}\label{fig2}
	\end{minipage}
\end{figure}

 Further, one can write down the evolution equation (\ref{13}) as the evolution of Hubble parameter with the scale factor as                                  \begin{equation}
	\frac{d H}{d a}+\frac{3}{2}(1+\omega)\frac{H}{a}=\frac{3}{2}\alpha
	(1+\omega)\frac{1}{a},    \label{32}      
\end{equation}
which on integration gives 
\begin{equation}
	H=\alpha -\delta(1+z)^{\frac{3}{2} (1+\omega)}    ,     \label{33}
\end{equation}
where $z$ is the amount of cosmological red shift $ \left(z=\dfrac{1}{a}-1  \right )$. Now introducing the dimensionless density parameter,  $ \Omega =\dfrac{\rho}{\rho _c}  ~\mbox{with}~ \rho_c =\dfrac{3 H^2}{8 \pi G}, \mbox{the critical density} $, the above equation can be written as 
\begin{equation}
	\frac{H^2}{H_0 ^2}=\Omega_{ \Lambda_0 }+\Omega _M (1+z)^{3(1+\omega)}+\Omega _{MP}(1+z)^{3(1+\omega _{MP})}    \label{34}
\end{equation}
where $\Omega_{ \Lambda_0 }= \left(1+{\dfrac{\delta}{H_0}}\right)^2$, $\Omega _M=\left(\dfrac{\delta}{H_0}\right)^2$, $\Omega _{MP}=2\dfrac{\delta}{H_0}\left(1+\dfrac{\delta}{H_0}\right)$ and $\omega_ {MP}=\left(\frac{\omega -1}{2}\right)$ with $\Omega_{ \Lambda_0 }+\Omega _M+ \Omega_{MP} =1 $.
From equation (\ref{34}) one can see that as $z\rightarrow -1$ i.e. $a\rightarrow\infty $, the present model approaches $\Lambda _{CDM}$ model. The evolution of the instantaneous equilibrium temperature of a system under non-equilibrium thermodynamic prescription can be written as \cite{Chakraborty:2014oya} ,
\begin{equation}
	\frac{\dot{T}}{T}+\omega\left(3H-\Gamma _d\right)=0.       \label{35}
\end{equation}
In the emergent scenario, one has (integrating equation (\ref{35}))  
\begin{equation}
	T= T_0 (1+z)^{3\omega} e^{\beta \omega (t-t_0)}    \label{36}
\end{equation}
where $T_0$ is the present measured value of temperature (at $t=t_0$) and $\beta$ is a constant. So, equations (\ref{34}) and (\ref{36}) represent the Hubble parameter and temperature respectively in terms of today's measured value.    The time evolution of $a$, $H$, $\phi$ and $\Gamma_d$ has been exhibited graphically in figure  \ref{fig1}. \textbf{Also the variation of thermodynamic parameters namely energy density $\rho$ and temperature $T$ with time $(t)$ and with equation of state parameter ( $\omega $) of the cosmic fluid have been shown in a $3d$ plot in figure  \ref{fig2}.} 
\section*{Discussion} The present work is an attempt to examine  whether emergent scenario of the Universe is possible under diffusive process. Considering kinetic model of the diffusion process, cosmological scalar field is chosen linearly to the Hubble parameter to obtain emergent scenario of the cosmic evolution \textbf{ and their variations with respect to time and equation of state parameter have been shown graphically ($3$d plot) in figure  \ref{fig2}.}
    Different thermodynamic parameters like energy density and temperature also have been determined under emergent scenario.\par Further it has been  established that such scalar field diffusion process corresponds to the particle creation mechanism \cite{Chakraborty:2014ora,Chakraborty:2014oya} in the non-equilibrium thermodynamic description. It is interesting to note that, for the non-singular particle creation process, [see  equation (\ref{31})] the barotropic index of the fluid can be restricted to $\omega < 0$ in the present scenario. \textbf{ Also for non-singular solution, the cosmological scalar field is chosen phenomenologically as proportional to the Hubble parameter and the  proportionality constant is found to be positive. } Finally, this work establishes  that the dissipative processes like diffusion, particle creation etc. may correspond to the evolution pattern of the universe as per the present observation. For future works, it may be attempted to find the Lagrangian formulation of such non-equilibrium thermodynamic phenomena to study the microscopic behaviour of the universe. 

\section*{Acknowledgements}  The author SM acknowledges UGC for
awarding Research fellowship and SC thanks Science and Engineering Research Board (SERB),India for awarding MATRICS Research grant support (File no. MTR/2017/000407).

\end{document}